\documentclass[10pt]{iopartmod}
\usepackage{amssymb}
\usepackage{amsmath}
\usepackage{makeidx}
\usepackage[dvips]{graphicx}

\input{tcilatex}

\begin{document}

\title{Slow wave resonance in periodic stacks of anisotropic layers}
\author{Alex Figotin and Ilya Vitebskiy}

\begin{abstract}
We consider a Fabry-Perot resonance (a transmission band edge resonance) in
periodic layered structures involving birefringent layers. In our previous
publication \cite{PRE05} we have shown that the presence of birefringent
layers with misaligned in-plane anisotropy can dramatically enhance the
performance of the photonic-crystal resonator. It allows to reduce its size
by an order of magnitude without compromising on its performance. The key
characteristic of the enhanced slow-wave resonator is that the Bloch
dispersion relation $\omega\left( k\right) $ of the periodic structure
displays a degenerate photonic band edge, in the vicinity of which the
dispersion curve can be approximated as $\Delta\omega\sim\left( \Delta
k\right) ^{4}$, rather than $\Delta\omega\sim\left( \Delta k\right) ^{2}$.
Such a situation can be realized in specially arranged stacks of misaligned
anisotropic layers. On the down side, the presence of birefringent layers
results in the slow wave resonance being coupled only with one (elliptic)
polarization component of the incident wave, while the other polarization
component is reflected back to space. In this paper we show how a small
modification of the periodic layered array can solve the above fundamental
problem and provide a perfect impedance match regardless of the incident
wave polarization, while preserving the giant slow-wave resonance
characteristic of a degenerate photonic band edge. Both features are of
critical importance for many practical applications, such as the enhancement
of various light-matter interactions, light amplification and lasing,
optical and microwave filters, antennas, etc.
\end{abstract}

\maketitle

\section{Introduction}

\begin{figure}[tbph]
\scalebox{0.8}{\includegraphics[viewport=-100 0 500 250,clip]{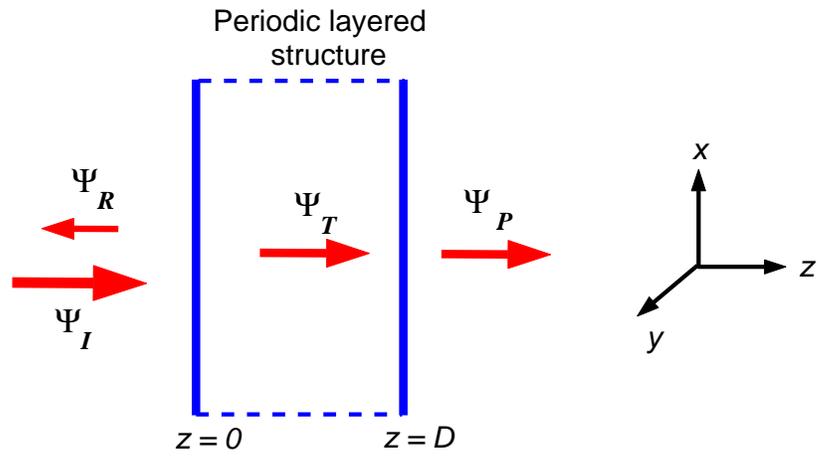}}
\caption{Scattering problem of a plane wave normally incident on a periodic
stack of dielectric layers. The indices $I$, $R$, and $P$ denote the
incident, reflected and transmitted waves, respectively. The field inside
the periodic medium is $\Psi_{T}$. In the case of a slow wave resonance, the
incident wave frequency lies in a transmission band of the periodic
structure, close to a band edge, as illustrated in Fig. \protect\ref{ISO}.}
\label{FSn}
\end{figure}

\begin{figure}[tbph]
\scalebox{0.8}{\includegraphics[viewport=0 0 500 400,clip]{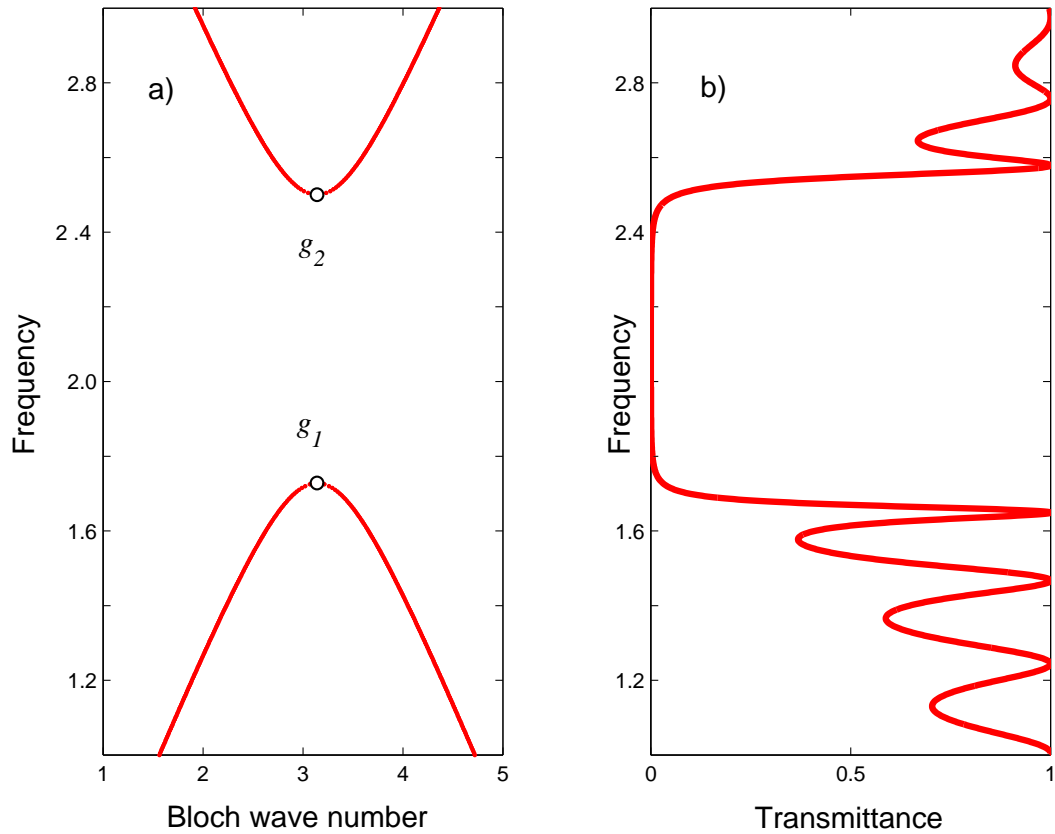}}
\caption{(a) A fragment of a typical Bloch $k-\protect\omega$ diagram of a
periodic array composed of non-birefringent layers; $g_{1}$ and $g_{2}$ are
the edges of the lowest photonic band gap. Each spectral branch is doubly
degenerate with respect to the wave polarization. (b) Transmission
dispersion $t\left( \protect\omega\right) $ of the respective finite
periodic stack; the sharp peaks near the edges of the transmission bands are
associated with slow-wave Fabry-Perot resonances, also known as transmission
band edge resonances. The location (\protect\ref{w_m}) of the resonance
peaks depends on the number $N$ of unit cells $L$ in the periodic stack.}
\label{ISO}
\end{figure}

Wave propagation in spatially periodic media, such as photonic crystals, can
be qualitatively different from any uniform substance. The differences are
particularly pronounced when the wavelength is comparable to the primitive
translation $L$ of the periodic structure \cite%
{Brill,LLEM,Joann,Yariv,Yeh,Chew,Notomi}. The effects of strong spatial
dispersion culminate at stationary points $\omega_{s}=\omega\left(
k_{s}\right) $ of the Bloch dispersion relation where the group velocity $%
u=\partial\omega/\partial k$ of a traveling Bloch wave vanishes 
\begin{equation}
\frac{\partial\omega}{\partial k}=0,\ \text{at }k=k_{s},~\omega=\omega
_{s}=\omega\left( k_{s}\right) .   \label{SP}
\end{equation}
One reason for this is that vanishing group velocity always implies a
dramatic increase in density of modes at the respective frequency. In
addition, vanishing group velocity also implies certain qualitative changes
in the eigenmode structure, which can be accompanied by some spectacular
effects in wave propagation. A particular example of the kind is the frozen
mode regime associated with a dramatic amplitude enhancement of the wave
transmitted to the periodic medium \cite%
{PRB03,PRE03,PRE05B,JMMM06,WRM06,PRE06}. In this paper, we focus on a
different slow-wave effect, namely, on a Fabry-Perot resonance in bounded
photonic crystals. This slow wave phenomenon, illustrated in Figs. \ref{FSn}
and \ref{ISO}, is also referred to as the transmission band edge resonance.
There are some similarities between the frozen mode regime and the slow-wave
resonance in plane-parallel photonic crystals. Both effects are associated
with vanishing group velocity at stationary point (\ref{SP}) of the Bloch
dispersion relation. As a consequence, both effects are strongly dependent
on specific type of spectral singularity (\ref{SP}). A fundamental
difference though is that the frozen mode regime is not a resonance
phenomenon in a sense that it is not particularly sensitive to the shape and
size of the photonic crystal. For instance, the frozen mode regime can occur
even in a semi-infinite periodic structure, where the incident plane wave is
converted to a frozen mode slowly propagating through the periodic medium
until it is absorbed \cite{PRB03,PRE03,PRE05B,JMMM06,WRM06,PRE06}. By
contrast, in the case of a slow wave resonance, the entire bounded periodic
structure acts as a resonator, resulting in a strong sensitivity of the
resonance behavior to the size and shape of the photonic crystal.

It is also important to distinguish between two qualitatively different
classes of photonic-crystal resonators. The first class comprises resonance
cavities where the role of periodic dielectric structure reduces to
electromagnetic (EM) field confinement by reflecting it back to the cavity
interior. The resonance frequency (or frequencies) of such photonic cavities
usually lies in a frequency gap (a stop-band) of the photonic crystal. The
periodic dielectric array here plays the role of a distributed Bragg
reflector. The number of resonance modes depends on the cavity size. It can
be a single mode localized on an isolated defect inside the photonic crystal 
\cite{Gor98,Lis06}. Or the cavity can support multiple resonances, if its
size significantly exceeds the light wavelength. More detailed information
on photonic crystal cavities can be found in numerous papers and monographs
on optics and photonics (see, for example, \cite{RCavity 95} and references
therein). In this paper, we will not further discuss this subject.

The second class of photonic-crystal resonators comprises the, so-called,
slow-wave resonators. They are qualitatively different from the band-gap
cavities. In slow-wave photonic-crystal resonators, the reflectors may not
be needed at all, as shown in the example in Fig. \ref{FSn}. The role of the
periodic structure here is to support slow EM waves. The resonance
frequencies lie in the transmission bands of the photonic crystal -- not in
band gaps. Although, the resonance frequencies can be very close to a band
edge (\ref{SP}), as shown in Fig. \ref{ISO}. A typical example of slow wave
resonance in a photonic crystal is presented by the transmission band edge
resonance, illustrated in Figs. \ref{FSn} and \ref{ISO}. In certain cases,
slow-wave resonators can provide significant advantages over cavity
resonators. They are used for the enhancement of light-matter interactions,
such as nonlinear and nonreciprocal effects, optical activity, light
amplification and lasing, etc. They can also be used in optical and
microwave filters, delay lines, as well as for the enhancement of antenna
gain and directionality. More detailed information can be found in an
extensive literature on the subject (see, for example, \cite{SWR
94,Yariv,Yeh,Chew,SL Scal1,SL Scal2,SL Joann,CR Yariv04,OSU DBE}, and
references therein).

In this paper we describe a slow-wave photonic-crystal resonator with
drastically reduced dimensions and enhanced performance, compared to that of
a common Fabry-Perot resonator based on a periodic stack of non-birefringent
layers. The idea is to employ periodic structures supporting the dispersion
relations different from those allowed in periodic arrays of
non-birefringent layers. Indeed, periodic arrays involving birefringent
layers can display stationary points (\ref{SP}) different from a regular
photonic band edge in Fig. \ref{ISO}(a). Some examples are shown in Fig. \ref%
{DR2}. Slow waves associated with such stationary points can produce giant
transmission band-edge resonances, much more powerful compared to those
achievable in common layered structures. The first step in this direction
was made in \cite{PRE05}, where it was shown that the transmission band-edge
resonance in the vicinity of a degenerate photonic band edge (DBE) in Fig. %
\ref{DR2}(b) produces much better results, compared to a regular photonic
band edge (RBE) of Figs. \ref{ISO}(a) and \ref{DR2}(a). Specifically, at the
frequency of DBE related giant slow-wave resonance, the electromagnetic
energy density inside the photonic-crystal can be estimated as%
\begin{equation}
\left\langle W_{DBE}\right\rangle \propto W_{I}N^{4},   \label{<W> DBE}
\end{equation}
where $W_{I}$ is the energy density of the incident wave and $N$ is the
total number of unit cells in the periodic stack. By comparison, the average
EM energy density at a regular transmission band-edge resonance in Fig. \ref%
{ISO} is%
\begin{equation}
\left\langle W_{RBE}\right\rangle \propto W_{I}N^{2}.   \label{<W> RBE}
\end{equation}
The estimations (\ref{<W> DBE}) and (\ref{<W> RBE}) imply that the Q-factor
of a DBE based slow-wave resonator can be by factor $N^{2}$ higher compared
to that of a RBE related Fabry-Perot resonator of the same size. And this a
huge difference! A detailed comparative analysis of the giant DBE related
slow wave resonance versus the regular transmission band edge resonance can
be found in \cite{PRE05}.

On the down side, periodic structures with birefringent layers have a
fundamental problem -- their reflectance and transmittance are essentially
dependent on the incident wave polarization. This dependence is particularly
strong near the edges of transmission bands, where the slow-wave resonances
occur. In particular, a DBE related giant transmission resonance described
in \cite{PRE05} is coupled only with one (elliptic) polarization component
of the incident wave, while the other polarization component is reflected
back to space by the photonic crystal boundary \cite{PRE05}. In other words,
at the resonance frequency, a periodic stack involving birefringent layers
acts as a polarizer, reflecting back to space roughly half of the incident
wave energy. This behavior is illustrated in Figs. \ref{tf_DBE} and \ref%
{AMnad18DBE}. Similar problem exists in all different modifications of the
frozen mode regime considered in \cite{PRB03,PRE03,PRE05B,JMMM06,WRM06,PRE06}%
. For many applications, such a polarization selectivity may not be
acceptable. In this paper we offer a solution to the above problem. We show
how to utilize all the incident wave energy, while preserving the
extraordinary performance of the DBE based photonic-crystal resonator. The
idea is to modify the periodic layered array so that instead of a degenerate
band edge, the respective dispersion curve develops a split photonic band
edge (SBE) shown in Fig. \ref{DR2}(b). Under certain conditions specified
below, the photonic resonator with a SBE will display a giant transmission
band edge resonance, similar to that of a DBE. But, in addition, the SBE\
resonator couples with the incident wave regardless of its polarization and,
therefore, utilizes all the incident EM radiation -- not just one
polarization component. The latter feature is of critical importance for a
variety of practical applications.

Similar approach can be applied not only to a photonic-crystal cavity
resonance, but also to all different modifications of the frozen mode regime
described in \cite{PRB03,PRE03,JMMM06,WRM06,PRE06}.

\section{Dispersion relation and energy flux in periodic arrays of
birefringent layers}

In this section we introduce some basic definitions and notations of
electrodynamics of stratified media composed of birefringent layers. A
detailed and consistent description of the subject, along with numerous
references, can be found in \cite{PRE05,PRB03,PRE03,PRE06}, where similar
notations and terminology are used. For simplicity, we restrict ourselves to
the case of a plane monochromatic wave normally incident on a layered
structure, as shown in Fig. \ref{FSn}. The results can be easily generalized
to the case of oblique incidence, as it was done, for example, in \cite%
{PRE03,PRE06}.

\subsection{Transverse waves in stratified media}

Time-harmonic electromagnetic field inside and outside the layered medium
can be described by the vector-column%
\begin{equation}
\Psi\left( z\right) =\left[ 
\begin{array}{c}
E_{x}\left( z\right) \\ 
E_{y}\left( z\right) \\ 
H_{x}\left( z\right) \\ 
H_{y}\left( z\right)%
\end{array}
\right] ,   \label{Psi}
\end{equation}
where $\vec{E}\left( z\right) $ and $\vec{H}\left( z\right) $ are
time-harmonic electric and magnetic fields. The $z$ direction is normal to
the layers. The values of $\Psi$ at any two different locations $z$ and $%
z^{\prime}$ are related by the transfer matrix $T\left( z,z^{\prime}\right) $
defined by%
\begin{equation}
\Psi\left( z\right) =T\left( z,z^{\prime}\right) \Psi\left( z^{\prime
}\right) .   \label{T}
\end{equation}
The elements of the transfer matrix are expressed in terms of material
tensors and other physical characteristics of the stratified medium.

Let $\Psi_{I}$, $\Psi_{R}$, and $\Psi_{P}$ be the incident, reflected, and
transmitted waves, respectively, as shown in Fig. \ref{FSn} . To the left of
the stack (at $z<0$), the electromagnetic field is a superposition of the
incident and reflected waves. To the right of the stack (at $z>D$), there is
only the transmitted wave. The field inside the periodic medium is denoted
as $\Psi_{T}$. Since all four transverse field components in (\ref{Psi}) are
continuous functions of $z$, we have the following boundary conditions at $%
z=0$ and $z=D$ in Fig. \ref{FSn}%
\begin{equation}
\Psi_{I}\left( 0\right) +\Psi_{R}\left( 0\right) =\Psi_{T}\left( 0\right) , 
\label{BC 0}
\end{equation}%
\begin{equation}
\Psi_{P}\left( D\right) =\Psi_{T}\left( D\right) .   \label{BC D}
\end{equation}

\subsection{Time-harmonic field inside periodic layered structure}

At any given frequency $\omega$, the time-harmonic field $\Psi_{T}\left(
z\right) $ inside the periodic stratified medium can be represented as a
superposition of Bloch eigenmodes, each of which satisfies the following
relation%
\begin{equation}
\Psi_{k}\left( z+L\right) =e^{ikL}\Psi_{k}\left( z\right) ,   \label{BW}
\end{equation}
or, equivalently,%
\begin{equation}
\Psi_{k}\left( z\right) =e^{ikz}\psi_{k}\left( z\right) ,\ \psi_{k}\left(
z+L\right) =\psi_{k}\left( z\right) .   \label{BM}
\end{equation}
The Bloch wave number $k$ is defined up to a multiple of $2\pi/L$. The
correspondence between $\omega$ and $k$ is referred to as the Bloch
dispersion relation. Real $k$ correspond to propagating (traveling) Bloch
modes. Propagating modes belong to different spectral branches $\omega\left(
k\right) $ separated by frequency gaps. In reciprocal and/or centrosymmetric
periodic structures, the Bloch dispersion relation is always symmetric with
respect to the points $k=0$ and $k=\pi/L$ of the Brillouin zone%
\begin{equation}
\omega\left( k_{0}+k\right) =\omega\left( k_{0}-k\right) , 
\label{w(k)=w(-k)}
\end{equation}
where%
\begin{equation}
k_{0}=0,\ \pi/L.   \label{k_0}
\end{equation}

In periodic structures composed of non-birefringent layers, every Bloch wave
is doubly degenerate with respect to polarization. Typical $k-\omega$
diagram for such a case is shown in Fig. \ref{ISO}(a). If, on the other
hand, some of the layers display an in-plane anisotropy or gyrotropy, the
polarization degeneracy can be lifted. The respective $k-\omega$ diagrams
are shown in Fig. \ref{DR2}.

Finally, if some of the layers are magnetically polarized and, in addition,
the periodic array is non-centrosymmetric, the dispersion relation can also
develop spectral asymmetry%
\begin{equation}
\omega\left( k\right) \neq\omega\left( -k\right) .   \label{w(k)<>w(-k)}
\end{equation}
This effect can be significant if the magnetic layers display an appreciable
nonreciprocal circular birefringence (magnetic Faraday rotation). Examples
of periodic layered structures with asymmetric dispersion relation (\ref%
{w(k)<>w(-k)}) can be found in \cite{PRE01,PRB03}. Further in this paper we
only discuss non-magnetic structures.

The speed of a traveling wave in a periodic medium is determined by the
group velocity $u=\partial\omega/\partial k$. Normally, every spectral
branch $\omega\left( k\right) $ develops stationary points (\ref{SP}) where
the group velocity of the corresponding propagating mode vanishes. Usually,
such points are located at the center and at the boundary of the Brillouin
zone%
\begin{equation}
k_{s}=k_{0}=0,~\pi/L.   \label{k_s=k_0}
\end{equation}
This is always the case in periodic layered structures composed of
non-birefringent layers, where all stationary points coincide with photonic
band edges, as shown in Fig. \ref{ISO}(a). If, on the other hand, some of
the layers in a unit cell are birefringent, then in addition to (\ref%
{k_s=k_0}), some dispersion curves can also develop a reciprocal pair of
stationary points corresponding to a general value of the Bloch wave number $%
k$, as shown in Fig. \ref{DR2}(b). The respective portion of the $k-\omega$
diagram can be described as a split band edge (SBE). The dispersion relation
can develop a DBE or a SBE only if the periodic layered array has
birefringent layers with misaligned in-plane anisotropy \cite{PRE05,PRE06}.
Example of such a layered structure is shown in Fig. \ref{StackAAB}.

Unlike propagating modes, evanescent Bloch modes are characterized by
complex wavenumbers $k=k^{\prime}+ik^{\prime\prime}$. Under normal
circumstances, evanescent modes decay exponentially with the distance from
the periodic structure boundaries. In such cases, the evanescent
contribution to $\Psi_{T}$ can be significant only in close proximity of the
surface or some other defects of the periodic structure. The situation can
change dramatically in the vicinity of a stationary point (\ref{SP}) of the
dispersion relation. At first sight, stationary points (\ref{SP}) relate
only to propagating Bloch modes. But in fact, in the vicinity of every
stationary point frequency $\omega_{s}$, the imaginary part $k^{\prime\prime}
$ of the Bloch wavenumber of at least one of the evanescent modes also
vanishes. As a consequence, the respective evanescent mode decays very
slowly, and its role may extend far beyond the photonic crystal boundary. In
addition, in some special cases, the electromagnetic field distribution in
the coexisting propagating and/or evanescent eigenmodes becomes very
similar, as $\omega$ approaches $\omega _{s}$. This can result in a
spectacular effect of coherent interference, such as the frozen mode regime 
\cite{PRB03,PRE03,WRM06,PRE06}. What exactly happens in the vicinity of a
particular stationary point essentially depends on its character and appears
to be very different for different types of singularity (\ref{SP}).

Using the transfer matrix (\ref{T}), the Bloch relation (\ref{BW}) can be
recast as%
\begin{equation}
T_{L}\Psi_{k}=e^{ikL}\Psi_{k}   \label{T BF}
\end{equation}
where $T_{L}$ is the transfer matrix of a unit cell%
\begin{equation*}
T_{L}=T\left( L,0\right) ,\ \Psi_{k}=\Psi_{k}\left( 0\right) . 
\end{equation*}

At any given frequency, there are four Bloch eigenmodes, propagating and/or
evanescent, with different polarizations and wave numbers%
\begin{equation}
\Psi_{k1}\left( z\right) ,\Psi_{k2}\left( z\right) ,\Psi_{k3}\left( z\right)
,\Psi_{k4}\left( z\right) .   \label{4Psi k}
\end{equation}
Depending on the frequency $\omega$, the full set (\ref{4Psi k}) of Bloch
eigenmodes may include only propagating modes, only evanescent modes, or
both. In any event, the respective set $\left[ k_{1},k_{2},k_{3},k_{4}\right]
$ of four Bloch wave numbers satisfies the relation%
\begin{equation}
\left[ k_{1},k_{2},k_{3},k_{4}\right] =\left[ k_{1}^{\ast},k_{2}^{\ast
},k_{3}^{\ast},k_{4}^{\ast}\right] ,   \label{k=k*}
\end{equation}
which is a direct consequence of $J$ unitarity of the transfer matrix $T$
(see \cite{PRE06} and references therein).

Taking into account (\ref{k=k*}), one can distinguish the following three
possibilities.

\begin{enumerate}
\item All four Bloch modes in (\ref{4Psi k}) are propagating%
\begin{equation}
k_{1}=k_{1}^{\ast},\ k_{2}=k_{2}^{\ast},\ k_{3}=k_{3}^{\ast},\
k_{4}=k_{4}^{\ast}   \label{4pr}
\end{equation}
In this case, two of the propagating modes have positive group velocities,
they are referred to as forward waves. The other two Bloch waves have
negative group velocities, they are referred to as backward waves.

\item All four Bloch modes in (\ref{4Psi k}) are evanescent%
\begin{equation}
k_{1}\neq k_{1}^{\ast},\ k_{2}\neq k_{2}^{\ast},\ k_{3}\neq k_{3}^{\ast },\
k_{4}\neq k_{4}^{\ast}   \label{4ev}
\end{equation}
This is the case when the frequency $\omega$ falls into a photonic band gap.
According to (\ref{k=k*}), one can assume that%
\begin{equation}
k_{1}=k_{2}^{\ast},\ k_{3}=k_{4}^{\ast}.   \label{4_ev}
\end{equation}
Two of the evanescent modes have $k^{\prime\prime}>0$; they are referred to
as forward evanescent modes. The other two evanescent modes have $k^{\prime
\prime}<0$; they are referred to as backward evanescent modes.

\item Two of the Bloch modes in (\ref{4Psi k}) are propagating modes, while
the other two are evanescent.%
\begin{equation}
k_{1}=k_{1}^{\ast},\ k_{2}=k_{2}^{\ast},\ k_{3}\neq k_{3}^{\ast},\ k_{4}\neq
k_{4}^{\ast}   \label{2pr 2ev}
\end{equation}
Again, according to (\ref{k=k*}), one can assume that%
\begin{equation}
k_{1}=k_{1}^{\ast},\ k_{2}=k_{2}^{\ast},\ k_{3}=k_{4}^{\ast}. 
\label{2_pr 2_ev}
\end{equation}
\end{enumerate}

In all cases, propagating modes with $u>0$ and evanescent modes with $%
k^{\prime\prime}>0$ are referred to as \emph{forward} waves. The propagating
modes with $u<0$ and evanescent modes with $k^{\prime\prime}<0$ are referred
to as \emph{backward} waves.

\subsection{EM energy flux in layered media}

The real-valued energy flux (the Poynting vector) associated with
time-harmonic field (\ref{Psi}) is%
\begin{equation}
S=\left[ \func{Re}\vec{E}\left( z\right) \times\func{Re}\vec{H}\left(
z\right) \right] =\frac{1}{4}\left(
E_{x}^{\ast}H_{y}-E_{y}^{\ast}H_{x}+E_{x}H_{y}^{\ast}-E_{x}H_{y}^{\ast}%
\right) .   \label{S(E,H)}
\end{equation}
The expression (\ref{S(E,H)}) can also be recast in the following compact
form \cite{WRM06} 
\begin{equation}
S=\frac{1}{2}\left( \Psi,J\Psi\right) ,   \label{S(Psi)}
\end{equation}
where%
\begin{equation}
J=\left[ 
\begin{array}{cccc}
0 & 0 & 0 & 1 \\ 
0 & 0 & -1 & 0 \\ 
0 & -1 & 0 & 0 \\ 
1 & 0 & 0 & 0%
\end{array}
\right] =J^{-1}.   \label{J}
\end{equation}

\subsubsection{EM energy flux in layered media}

In the case of a lossless stratified medium, the Poynting vector $S$ in (\ref%
{S(Psi)}) is independent of the coordinate $z$.

In a periodic layered medium, the column vector $\Psi\left( z\right) $ in (%
\ref{S(Psi)}) can be represented as a linear combination of four Bloch
components (\ref{4Psi k})%
\begin{equation}
\Psi_{T}\left( z\right) =\Psi_{k1}\left( z\right) +\Psi_{k2}\left( z\right)
+\Psi_{k3}\left( z\right) +\Psi_{k4}\left( z\right) ,\ 0<z<D. 
\label{4Psi_k}
\end{equation}
Substituting (\ref{4Psi_k}) into (\ref{S(Psi)}) yields%
\begin{equation}
S=\frac{1}{2}\sum_{i,j=1}^{4}\left( \Psi_{i},J\Psi_{j}\right) . 
\label{S(4Psi)}
\end{equation}
At this point we can apply the orthogonality relation%
\begin{equation}
\left( \Psi_{i},J\Psi_{j}\right) =0\text{, if }k_{i}\neq k_{j}^{\ast}, 
\label{[i,j]=0}
\end{equation}
the proof of which can be found in \cite{WRM06}, P. 327. The expression (\ref%
{S(4Psi)}) together with (\ref{[i,j]=0}) lead to the following conclusions
regarding the energy flux of a time-harmonic electromagnetic field in
lossless periodic layered media.

\begin{enumerate}
\item[1)] The contribution of each propagating Bloch mode to the total
energy flux is independent of the presence or absence of other Bloch
eigenmodes with the same frequency%
\begin{equation}
S=\dsum \limits_{i=1}S_{i}=\frac{1}{2}\dsum \limits_{i=1}\left(
\Psi_{i},J\Psi_{i}\right) ,   \label{S(pr)}
\end{equation}
where the summation runs over all propagating eigenmodes. The number of
propagating modes can be 4, 2, or 0, depending on which of the cases (\ref%
{4pr}), (\ref{2pr 2ev}), or (\ref{4ev}) we are dealing with.

\item[2)] The contribution of evanescent eigenmodes to the energy flux
depends on how many of them exist at this particular frequency $\omega$.

\begin{enumerate}
\item In the case (\ref{2pr 2ev}) of two evanescent modes $\Psi_{3}$ and $%
\Psi_{4}$ we have%
\begin{equation}
S=\func{Re}\left( \Psi_{3},J\Psi_{4}\right) ,\text{ where }%
k_{4}=k_{3}^{\ast},   \label{S(2ev)}
\end{equation}
which implies that only a pair of evanescent modes with conjugate wave
numbers can contribute to the energy flux. The respective contribution (\ref%
{S(2ev)}) is independent of the presence of propagating modes $\Psi_{1}$ and 
$\Psi_{2}$. In accordance with Eq. (\ref{S(2ev)}), a single evanescent mode,
either $\Psi_{3}$ or $\Psi_{4}$, does not produce energy flux on its own.

\item In the case (\ref{4ev}) of four evanescent modes we have%
\begin{equation}
S=\func{Re}\left( \Psi_{1},J\Psi_{2}\right) +\func{Re}\left(
\Psi_{3},J\Psi_{4}\right) ,\text{ where }k_{2}=k_{1}^{\ast},k_{4}=k_{3}^{%
\ast},   \label{S(4ev)}
\end{equation}
which implies that either of the two pairs of evanescent modes with
conjugate wave numbers contribute to the energy flux independently of each
other.
\end{enumerate}
\end{enumerate}

In the general case of the time-harmonic EM field $\Psi_{T}\left( z\right) $
being a superposition (\ref{4Psi_k}) of several Bloch modes, propagating and
evanescent, the contribution of each propagating mode $\Psi_{k}$ to the
total energy flux can be expressed in terms of its group velocity and
amplitude%
\begin{equation}
S_{k}=W_{k}u_{k},   \label{S=Wu}
\end{equation}
where%
\begin{equation}
W_{k}\propto\left\langle \left\vert \Psi_{k}\right\vert \right\rangle
^{2}=\left\langle \left\vert \psi_{k}\right\vert \right\rangle ^{2} 
\label{Wk}
\end{equation}
In the particular case of a single propagating mode, the quantity $W_{k}$ in
(\ref{S=Wu}) is equal to the energy density averaged over a unit cell $L$.

By contrast, a single evanescent mode never transfers the energy. Only the
combination (\ref{S(2ev)}) or (\ref{S(4ev)}) of forward and backward
evanescent modes can contribute to the energy flux.

\subsubsection{Transmission/reflection coefficients}

Let $S_{I}$, $S_{R}$, $S_{T}$, and $S_{P}$ be the energy fluxes of the
respective waves in Fig. \ref{FSn}. The transmission/reflection coefficients 
$t$ and $r$ are defined as%
\begin{equation}
t=\frac{S_{P}}{S_{I}},\ r=-\frac{S_{R}}{S_{I}}.   \label{tn, rn}
\end{equation}
In the case of losses medium, the Poynting vector $S$ is independent of $z$%
\begin{equation*}
S\equiv S_{I}+S_{R}=S_{T}=S_{P}. 
\end{equation*}
In such a case%
\begin{equation}
t=1-r=S/S_{I}.   \label{tn=1-rn}
\end{equation}
Otherwise,%
\begin{equation*}
t+r+a=1, 
\end{equation*}
where $a>0$ is the absorption coefficient of the stack.

\subsection{Single mode excitation regime}

To explain the nature of strong polarization dependence of the transmission
band edge resonance in periodic stacks of anisotropic layers we introduce
some basic concepts related to semi-infinite periodic layered structures.
Then, we will see how those concepts apply to finite periodic stacks.

Consider a plane monochromatic wave incident on a semi-infinite periodic
structure with the same unite cell $L$ as that of the finite periodic stack
in Fig. \ref{FSn}. In other words, let us assume that the right-hand
boundary of the photonic crystal in Fig. \ref{FSn} is absent. In such a
case, the EM field $\Psi_{T}\left( z\right) $ inside the periodic medium is
a superposition of two forward Bloch waves, each of which can be either
propagating with $u>0$, or evanescent with $k^{\prime\prime}>0$. The
backward Bloch modes (propagating with $u<0$ and evanescent with $%
k^{\prime\prime}<0$) do not contribute to $\Psi_{T}\left( z\right) $ in the
case of a semi-infinite periodic structure. According to \cite{PRE06}, the
incident wave polarization can always be chosen so that only one of the two
forward Bloch waves is excited. Such a situation is referred to as a \emph{%
single mode excitation }regime. The respective two polarizations of the
incident wave are referred to as the \emph{single mode excitation }%
polarizations. In periodic structures involving birefringent layers, a
single mode excitation polarization is, generally, elliptic. Note, that if
all the layers are non-birefringent, then, due to polarization degeneracy,
the very concept of a single mode excitation regime at normal propagation
makes no sense. In the case of oblique incidence, though, a single mode
excitation regime exists even in periodic structures composed of
non-birefringent layers; the respective incident wave polarizations are
linear and referred to as\ TE and TM.

As we already mentioned, the dispersion relation of a periodic structure
involving birefringent layers usually does not display polarization
degeneracy. In particular, the periodic stack in Fig. \ref{StackAAB} can
display polarization degeneracy at normal propagation only if the
misalignment angle (\ref{phi}) between the adjacent anisotropic layers\ $%
A_{1}$ and $A_{2}$ is $\pi/2$. In all numerical examples in this paper, the
misalignment angle $\phi$ is different from $0$ and $\pi/2$, because
otherwise the existence of a DBE and a SBE would be ruled out by symmetry 
\cite{PRE06}. Therefore, none of the spectral branches in Fig. \ref{DR2} is
degenerate with respect to polarization.

\subsubsection{The vicinity of a RBE}

Consider the vicinity of a RBE $g_{1}$ or $g_{2}$ in Fig. \ref{DR2}(a). If
the incident wave frequency lies in a transmission band close enough to the
respective RBE, the transmitted wave $\Psi_{T}\left( z\right) $ inside the
semi-infinite periodic layered structure is a superposition of two forward
Bloch waves%
\begin{equation}
\Psi_{T}\left( z\right) =\Psi_{pr}\left( z\right) +\Psi_{ev}\left( z\right)
,\ \ \ \ z>0,   \label{pr + ev}
\end{equation}
one of which is propagating and the other is evanescent. There are two
single mode excitation regimes at such a frequency. The first one produces
only propagating mode $\Psi_{pr}\left( z\right) $ at $z>0$, while the other
produces only evanescent mode $\Psi_{ev}\left( z\right) $ at $z>0$. The
respective transmission coefficients satisfy the following relations%
\begin{equation}
\text{if }\Psi_{T}\left( z\right) =\Psi_{pr}\left( z\right) \text{, \ then }%
t=t_{pr}>0,   \label{t_pr}
\end{equation}%
\begin{equation}
\text{if }\Psi_{T}\left( z\right) =\Psi_{ev}\left( z\right) \text{, \ then }%
t=t_{ev}=0.   \label{t_ev}
\end{equation}
Eq. (\ref{t_ev}) reflects the fact that according to (\ref{S(2ev)}), a
single evanescent mode does not contribute to the energy flux.

The relations (\ref{t_pr}) and (\ref{t_ev}) imply that only one of the two
polarization components of the incident wave (the one producing the forward
propagating mode $\Psi_{pr}$) can be transmitted to the semi-infinite
photonic crystal. The other polarization component of the incident wave (the
one producing only evanescent mode $\Psi_{ev}$) will be totally reflected
back to space.

Now, what happens if the periodic structure has a finite thickness? In such
a case, the relation (\ref{pr + ev}) does not apply. Instead, the EM field $%
\Psi_{T}\left( z\right) $ inside the finite periodic stack is a
superposition of all four Bloch waves (\ref{4Psi k}), including both forward
and backward modes. If the number $N$ of unit cells in the periodic stack is
significant, then at certain frequencies, the pair of propagating modes (one
forward and one backward ) form a standing wave (\ref{1sw}) constituting the
transmission band edge resonance. An important point, though, is that such a
resonance is excited only by one (elliptic) polarization component of the
incident wave -- the same component that produces a single propagating mode
inside the semi-infinite structure. If, on the other hand, the incident wave
polarization is such that it produces a single evanescent mode inside the
semi-infinite structure, then it will be reflected back to space from the
finite periodic stack as well. In practice, the relation (\ref{t_ev}) is
valid even for the finite periodic structures composed of as few as several
unit cells $L$.

\subsubsection{The vicinity of a DBE}

Consider now the vicinity of a DBE $d$ in Fig. \ref{DR2}(b). If the incident
wave frequency lies in the upper transmission band close to the DBE $d$, the
transmitted wave $\Psi_{T}\left( z\right) $ inside the semi-infinite
periodic layered structure still is a superposition (\ref{pr + ev}) of one
propagating and one evanescent Bloch waves. But remarkably, in close
proximity of the DBE, the column vectors $\Psi_{pr}$ and $\Psi_{ev}$ in (\ref%
{pr + ev}) become nearly parallel to each other, while their amplitude
diverges \cite{PRE06}. More specifically, at the photonic crystal boundary
at $z=0$%
\begin{equation}
\Psi_{pr}\left( 0\right) \approx-\Psi_{ev}\left( 0\right) \propto \left\vert
\omega-\omega_{d}\right\vert ^{-1/4}\text{, as }\omega
\rightarrow\omega_{d},   \label{DI 4b}
\end{equation}
provided that the incident wave has some general polarization and fixed
amplitude. The destructive interference (\ref{DI 4b}) ensures that the
boundary condition (\ref{BC 0}) can be satisfied, while both Bloch
contributions to $\Psi_{T}\left( z\right) $ diverge. As the distance $z$
from the slab boundary increases, the evanescent component $\Psi_{ev}\left(
z\right) $ dies out%
\begin{equation}
\Psi_{ev}\left( z\right) \approx\Psi_{ev}\left( 0\right) \exp\left(
-zk^{\prime\prime}\right) ,   \label{Psi_ev DBE}
\end{equation}
while the propagating component $\Psi_{pr}\left( z\right) $ remains constant
and very large. Eventually, as the distance $z$ further increases, the
transmitted wave $\Psi_{T}\left( z\right) $ reaches its large saturation
value corresponding to its propagating component $\Psi_{pr}\left( z\right) $%
. This constitutes the DBE related frozen mode regime in a semi-infinite
photonic crystal \cite{PRE06}.

Generally, the incident wave polarization can always be chosen so that only
one of the two forward Bloch waves in (\ref{pr + ev}) is excited in the
semi-infinite photonic crystal (a single mode excitation regime). What
happens to the frozen mode regime in such a case? Obviously, the destructive
interference (\ref{DI 4b}) cannot occur in a single mode excitation regime.
As a consequence, the incident wave is reflected back to space without
producing the frozen mode, regardless of whether the single mode excitation
regime produces a single propagating or a single evanescent Bloch mode \cite%
{PRE06}. The latter circumstance is related to the fact that the incident
wave polarizations producing a single propagating mode and a single
evanescent mode become indistinguishable as $\omega\rightarrow\omega_{d}$.

The fact that all four Bloch modes (\ref{4Psi k}) are nearly parallel to
each other in the vicinity of a DBE is also very important for understanding
the DBE related giant transmission band edge resonance in a finite photonic
crystal. Indeed, as we pointed out earlier, in the case of RBE related
resonance, the incident wave producing a single propagating mode is coupled
with the resonance mode, while the incident wave producing a single
evanescent mode is not. By contrast, in the case of DBE related resonance,
both polarizations are nearly indistinguishable. As a result, neither of the
two single mode regime polarizations is coupled with the resonance mode.
Instead, the best result is achieved if the incident wave polarization is
orthogonal to that of single mode regime polarization. A numerical
illustration of this point is given in Fig. \ref{tf_DBE}.

In summary, the DBE related transmission band edge resonance is not excited
if the incident wave polarization corresponds to the single mode excitation
regime. This implies that if the incident wave polarization is randomly
chosen, then on average, at least half of the incident wave energy is
reflected back to space from the photonic crystal boundary, without
contributing to the resonance. The above problem can be overcome in slightly
modified periodic structures displaying a SBE, rather than a DBE.

\section{Simplified geometrical description of transmission band-edge
resonance}

Steady-state Fabry-Perot resonance in a pane-parallel photonic crystal is
commonly described as a standing wave composed of a pair of reciprocal
propagating Bloch waves (\ref{1sw}) with equal and opposite Bloch wave
numbers and vanishing group velocities. The nodes of the standing wave
coincide with the photonic crystal boundaries at $z=0$ and $z=D$. This
simple representation works very well in the vicinity of a regular photonic
band edge (RBE). But, in the cases of DBE or SBE related giant transmission
band edge resonance, the above simple picture should be significantly
modified. We start this section with a brief description of the standing
wave concept as applied to a regular transmission band edge resonance. Then
we proceed to a comparative analysis of giant transmission band edge
resonances in the vicinity of a DBE and SBE.

\subsection{Slow-wave resonance as a standing wave}

Consider a finite plane-parallel periodic structure composed of $N$ unit
cells $L$. If the number $N$ is significant, the electromagnetic properties
of the periodic stack can be effectively described in terms of the Bloch
eigenmodes (\ref{BM}) of the corresponding infinite periodic layered medium.
For instance, a periodic stack having as few as several periods $L$ can
display almost total reflectivity at photonic band gap frequencies, which
results from the absence of propagating Bloch modes at the respective
frequency range. A typical dependence of the finite stack transmittance on
the incident wave frequency is shown in Fig. \ref{ISO}(b). The sharp peaks
in transmission bands correspond to transmission band edge resonances. The
resonance frequencies are located close to a photonic band edge, where the
group velocity is very low. At each resonance, the electromagnetic field $%
\Psi_{T}\left( z\right) $ inside the periodic stack is close to a standing
wave composed of a pair of reciprocal Bloch waves with large and nearly
equal amplitudes and low group velocity%
\begin{equation}
\Psi_{T}\left( z\right) =\Psi_{k}\left( z\right) +\Psi_{-k}\left( z\right) , 
\label{1sw}
\end{equation}
where%
\begin{equation*}
\text{ }k=k_{m},\ \ 0<z<D=NL. 
\end{equation*}
The photonic crystal boundaries at $z=0$ and $z=D$ coincide with the
standing wave nodes, where the forward and backward Bloch components
interfere destructively to meet the boundary conditions (\ref{BC 0}) and (%
\ref{BC D}). The latter circumstance determines the wave numbers of the
Bloch eigenmodes contributing to the resonance field inside the periodic
structure%
\begin{equation}
\left( k_{m}-k_{0}\right) \approx\pm\frac{\pi}{NL}m,\ \ m=1,2,..., 
\label{k_m}
\end{equation}
The wave number $k_{0}$ is given in (\ref{k_0}). The integer$\ m$ denotes
the resonance peaks in order of their distance from the respective photonic
band edge. The respective resonance frequencies are expressed in terms of
the dispersion relation%
\begin{equation}
\omega_{m}=\omega\left( k_{m}\right) .   \label{w_m}
\end{equation}
According to (\ref{k_m}), the proximity of the resonances to the photonic
band edge is determined by the number $N$ of unit cells in the periodic
stack. The expressions (\ref{k_m}) and (\ref{w_m}) only apply if%
\begin{equation}
N\gg m.   \label{N>>1}
\end{equation}

Let us introduce the following dimensionless notations for the small
deviation of the wave number and the frequency from the respective
stationary point 
\begin{equation}
\kappa=\left\vert k-k_{0}\right\vert L,\ \ \nu\left( \kappa\right) =\left[
\omega\left( k\right) -\omega\left( k_{0}\right) \right] L/c. 
\label{kappa, nu}
\end{equation}
According to (\ref{k_m}) and (\ref{w_m}), the resonance values of $\kappa$
and $\nu$ are 
\begin{equation}
\kappa_{m}\approx\frac{\pi}{N}m\ll1,\ \nu_{m}\approx\nu\left( \kappa
_{m}\right) .   \label{kappa_m, nu_m}
\end{equation}
The most powerful resonance is usually the one closest to the respective
photonic band edge%
\begin{equation}
\ \kappa_{1}\approx\frac{\pi}{N}\ll1,\ \nu_{1}\approx\nu\left( \kappa
_{1}\right) .   \label{kappa_1, nu_1}
\end{equation}

\subsection{Energy density and energy flux at resonance frequency}

Assume that the amplitude of the incident wave in Fig. \ref{FSn} is unity.
Due to the boundary condition (\ref{BC 0}), we have at the left-hand
boundary of the photonic crystal in Fig. \ref{FSn}%
\begin{equation}
\Psi_{T}\left( 0\right) =\Psi_{k}\left( 0\right) +\Psi_{-k}\left( 0\right)
\propto1.   \label{Psi(0)=1}
\end{equation}
At resonance frequency, the transmission coefficient (\ref{tn=1-rn}) is
close to unity%
\begin{equation}
t_{m}=S/S_{I}\propto1.   \label{t_m=1}
\end{equation}
Therefore, the relation similar to (\ref{Psi(0)=1}) also applies at the
right-hand boundary of the periodic stack%
\begin{equation}
\Psi_{T}\left( D\right) =\Psi_{k}\left( D\right) +\Psi_{-k}\left( D\right)
\propto1.   \label{Psi(D)=1}
\end{equation}

Let us see under what conditions the finite resonance transmission (\ref%
{t_m=1}) is compatible with the vanishing group velocity in the vicinity of
stationary point (\ref{SP}), where the transmission band edge resonance
occurs. According to (\ref{S=Wu}), the energy flux associated with the
time-harmonic field $\Psi_{T}\left( z\right) $ in (\ref{1sw}) is 
\begin{equation}
S=S_{k}+S_{-k}\approx W_{k}u\left( k\right) +W_{-k}u\left( -k\right) =\left(
W_{k}-W_{-k}\right) u\left( \kappa\right) ,   \label{S 1sw}
\end{equation}
where%
\begin{equation}
\ \ u\left( \kappa\right) =\left\vert u\left( k\right) \right\vert
=\left\vert u\left( -k\right) \right\vert ,   \label{|u|}
\end{equation}
and%
\begin{equation*}
W_{k}\propto\left\langle \left\vert \Psi_{k}\right\vert ^{2}\right\rangle ,\
W_{-k}\propto\left\langle \left\vert \Psi_{-k}\right\vert ^{2}\right\rangle
. 
\end{equation*}
In the vicinity of stationary point (\ref{SP}), the magnitude $u\left(
\kappa\right) $ of the group velocity in (\ref{S 1sw}) is vanishingly small.
The fact that the resonance energy flux remains of the order of unity
implies that the amplitude of the Bloch components in (\ref{S 1sw})
increases so that%
\begin{equation}
W_{k}-W_{-k}\propto u^{-1},\text{ as }u\rightarrow0,   \label{W-W=1/u}
\end{equation}
or, equivalently,%
\begin{equation}
\left\langle \left\vert \Psi_{k}\right\vert ^{2}\right\rangle -\left\langle
\left\vert \Psi_{-k}\right\vert ^{2}\right\rangle \propto u^{-1},\text{ as }%
u\rightarrow0.   \label{Psi-Psi=1/u}
\end{equation}
Thus, finite transmittance at $u\rightarrow0$ requires the divergence (\ref%
{Psi-Psi=1/u})\ of the amplitude of the Bloch components.

The requirements (\ref{Psi(0)=1}) and (\ref{Psi(D)=1}) together with (\ref%
{Psi-Psi=1/u}) imply that although the amplitudes of the forward and
backward Bloch components both diverge as $u\rightarrow0$, they must
interfere destructively near the photonic crystal boundaries%
\begin{equation*}
\Psi_{k}\left( 0\right) \approx-\Psi_{-k}\left( 0\right) \propto u^{-1}, 
\end{equation*}
and%
\begin{equation*}
\Psi_{k}\left( D\right) \approx-\Psi_{-k}\left( D\right) \propto u^{-1}. 
\end{equation*}
In order to reconcile the boundary condition (\ref{Psi(0)=1}) with the
requirement (\ref{Psi-Psi=1/u}) of a finite energy flux, we have to impose
the following requirement on the amplitudes of the two Bloch components%
\begin{equation}
W_{k}-W_{-k}\propto\sqrt{W_{k}}\propto u^{-1},\text{ \ as \ }u\rightarrow0. 
\label{W-W}
\end{equation}

The relation (\ref{W-W}) was derived under the assumption that the
time-harmonic filed $\Psi_{T}\left( z\right) $ inside the periodic medium is
a superposition (\ref{1sw}) of one forward and one backward propagating
Bloch waves with equal and opposite wave numbers and group velocities. This
assumption is warranted in the vicinity of a regular photonic band edge of a
periodic stack with birefringent layers, such as those shown in Fig. \ref%
{DR2}(a). In a periodic stack of non-birefringent layers, at any given
frequency there are two pairs of reciprocal Bloch waves with equal wave
numbers but different polarizations. Still, because of the polarization
degeneracy, the above consideration can be directly applied to the latter
case, as well.

By contrast, in the DBE and SBE related transmission band edge resonances,
the contribution of all four Bloch components to the resonance field $\Psi
_{T}\left( z\right) $ is absolutely essential. In such cases, the treatment
of the transmission band edge resonance as a simple standing wave (\ref{S
1sw}) is not adequate.

\subsection{Regular photonic band edge (RBE)}

Let us apply the relation (\ref{W-W}) to the classical case of Fabry-Perot
resonance in the vicinity of a regular photonic band edge (RBE). Using the
notations (\ref{kappa, nu}), the dispersion relation in the vicinity of a
RBE can be approximated as%
\begin{equation}
\nu\left( \kappa\right) \approx\frac{a}{2}\kappa^{2}.   \label{DR RBE}
\end{equation}
The group velocity magnitude is%
\begin{equation}
u\approx a\kappa\approx\sqrt{2\nu/a},\ 0<\nu/a.   \label{u RBE}
\end{equation}
According to (\ref{kappa_m, nu_m}), the resonance values of the wave number
and the frequency are, respectively%
\begin{equation*}
\kappa_{m}=\frac{\pi m}{N},\ \nu_{m}\approx\frac{a}{2}\left( \frac{\pi m}{N}%
\right) ^{2}. 
\end{equation*}
The group velocity magnitude at resonance is%
\begin{equation}
u_{m}\approx a\frac{\pi m}{N}.   \label{u1 RBE}
\end{equation}
Substitution of $u_{m}$ to (\ref{W-W}) yields 
\begin{equation*}
W_{\kappa}-W_{-\kappa}\propto\sqrt{W_{\kappa}}\propto\frac{N}{am}. 
\end{equation*}
The energy density distribution $W\left( z\right) $ at RBE related
resonances is typical of a standing wave%
\begin{equation}
W\left( z\right) \propto W_{I}\left( \frac{N}{m}\right) ^{2}\sin ^{2}\left( 
\frac{\pi m}{NL}z\right) ,~~m=1,2,....   \label{W(z) RBE}
\end{equation}

In the case of the most powerful first transmission resonance ( the one with 
$m=1$), the two Bloch components in (\ref{1sw}) interfere constructively in
the middle of the periodic structure. The respective field intensity is%
\begin{equation}
W_{\kappa}\propto W_{I}N^{2}.   \label{W RBE}
\end{equation}

\subsection{Degenerate photonic band edge (DBE)}

The dispersion relation in the vicinity of a DBE $d$ in Fig. \ref{DR2}(b)
can be approximated as%
\begin{equation}
\nu\left( \kappa\right) \approx\frac{b}{4}\kappa^{4}.   \label{DR DBE}
\end{equation}
The magnitude $u$ of the group velocity of each of the reciprocal pair of
propagating modes is%
\begin{equation}
u\approx b\kappa^{3}\approx\sqrt{8}\left( b\nu^{3}\right) ^{1/4},\ 0<b\nu. 
\label{u DBE}
\end{equation}
Unlike the case of RBE related resonance, in the vicinity of a DBE, the EM
field inside the periodic medium cannot be approximated by a simple standing
wave (\ref{1sw}). Instead, we have to take into account the evanescent mode
contribution, which appears to be comparable to that of the propagating
modes \cite{PRE05}. As the result, the simple geometrical conditions (\ref%
{k_m}) and (\ref{kappa_m, nu_m}) give only rough order of magnitude estimate
of the resonance values of the wave number and frequency. More importantly,
the maximum field intensity inside the photonic crystal at the frequency of
the transmission band-edge resonance is now%
\begin{equation}
W_{k}\propto W_{I}N^{4},   \label{W DBE}
\end{equation}
which is by factor $N^{2}$ larger than that of a regular transmission
resonance given in (\ref{W RBE}).

The detailed analysis of the DBE related giant transmission resonance was
carried out in \cite{PRE05}. Here, we would like to highlight the down side
of this effect, which is the polarization selectivity illustrated in Figs. %
\ref{tf_DBE} and \ref{AMnad18DBE}. As we already stated before, the DBE
related giant transmission resonance is coupled only with one (elliptic)
polarization component of the incident wave, while the other polarization
component is reflected back to space by the photonic crystal boundary. The
reflected polarization component corresponds to the single mode excitation
regime, explained at the end of Section 2. A solution to this fundamental
problem is to slightly modify the periodic structure so that instead of a
DBE, the respective spectral branch displays a SBE. Example of a SBE is
shown in Fig. \ref{DR2}(b).

\section{Transmission band-edge resonance in the vicinity of a split
photonic band edge (SBE)}

\subsection{Simplified description of SBE related resonance}

In the previous section we described the resonance conditions for a single
pair (\ref{1sw}) of reciprocal Bloch waves. In periodic stacks involving
birefringent layers, a slow-wave resonance produced by a single pair of
reciprocal Bloch waves can be excited by only one polarization component of
the incident plane wave. The other polarization component is not coupled
with the respective resonance mode. This implies that if the incident wave
polarization is random, at least half of the incident wave energy will not
enter the periodic medium and will be reflected back to space by the
photonic crystal boundary.

\begin{figure}[tbph]
\scalebox{0.8}{\includegraphics[viewport=0 0 500 400,clip]{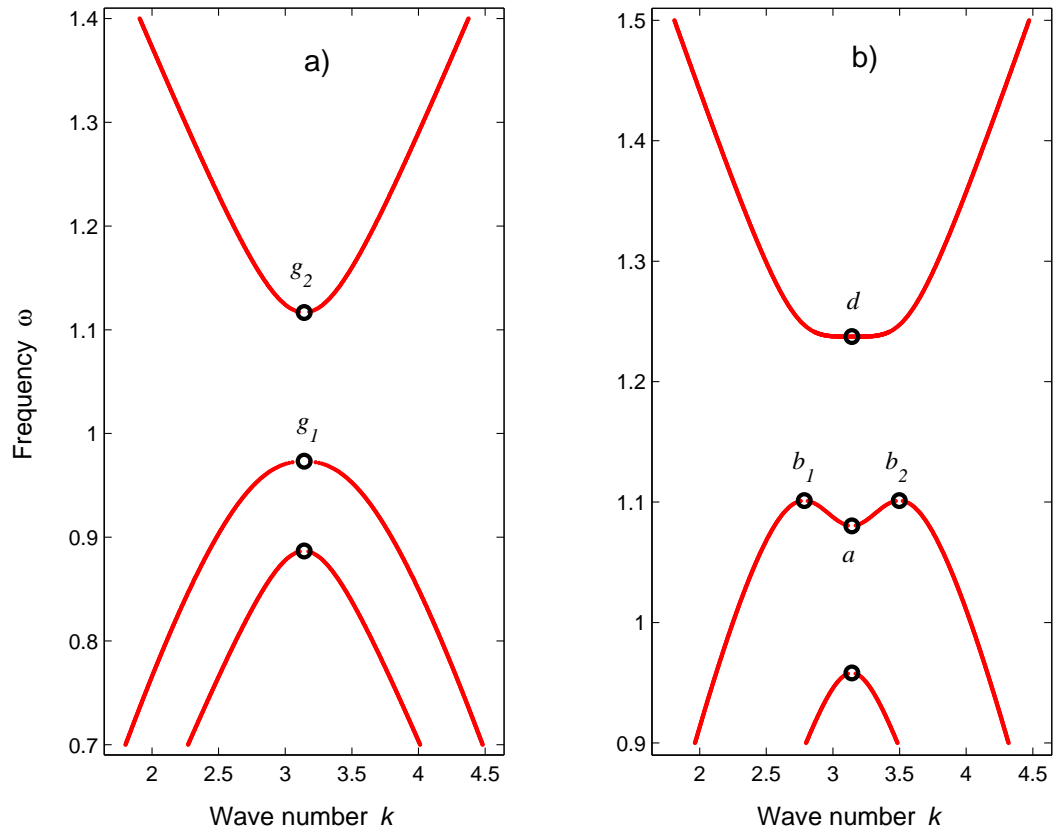}}
\caption{Fragments of the $k-\protect\omega$ diagrams of the periodic
layered structure in Fig. \protect\ref{StackAAB} for two different values of
the ratio $A/B$ of the layer thicknesses. Regular, degenerate , and split
photonic band edges are denoted by symbols $g$, $d$, and $b$, respectively.
The Bloch wave number $k$ and the frequency $\protect\omega$ are expressed
in units of $1/L$ and $c/L$.}
\label{DR2}
\end{figure}

\begin{figure}[tbph]
\scalebox{0.8}{\includegraphics[viewport=0 0 500 400,clip]{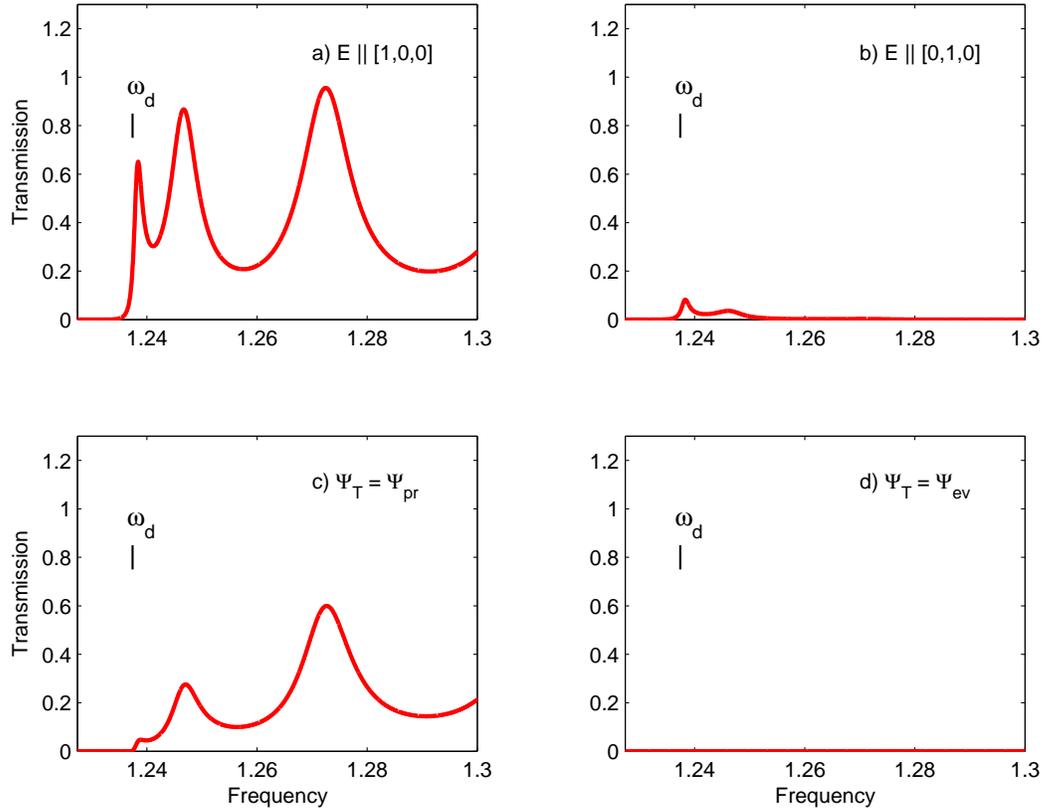}}
\caption{Transmission dispersion of the periodic stack of 18 unit cells at
frequency range including a DBE at $\protect\omega=\protect\omega_{d}$. The
respective $k-\protect\omega$ diagram is shown in Fig. \protect\ref{DR2}(b).
In the cases (a) and (b), the incident wave in linearly polarized. In the
cases (c) and (d), the incident wave polarization is adjusted so that at any
given frequency it corresponds to a single mode excitation regime: in the
case (c) it is a single propagating mode, while in the case (d) it is a
single evanescent mode. Obviously, in the latter case the incident wave is
reflected back to space. }
\label{tf_DBE}
\end{figure}

\begin{figure}[tbph]
\scalebox{0.8}{\includegraphics[viewport=0 0 500 400,clip]{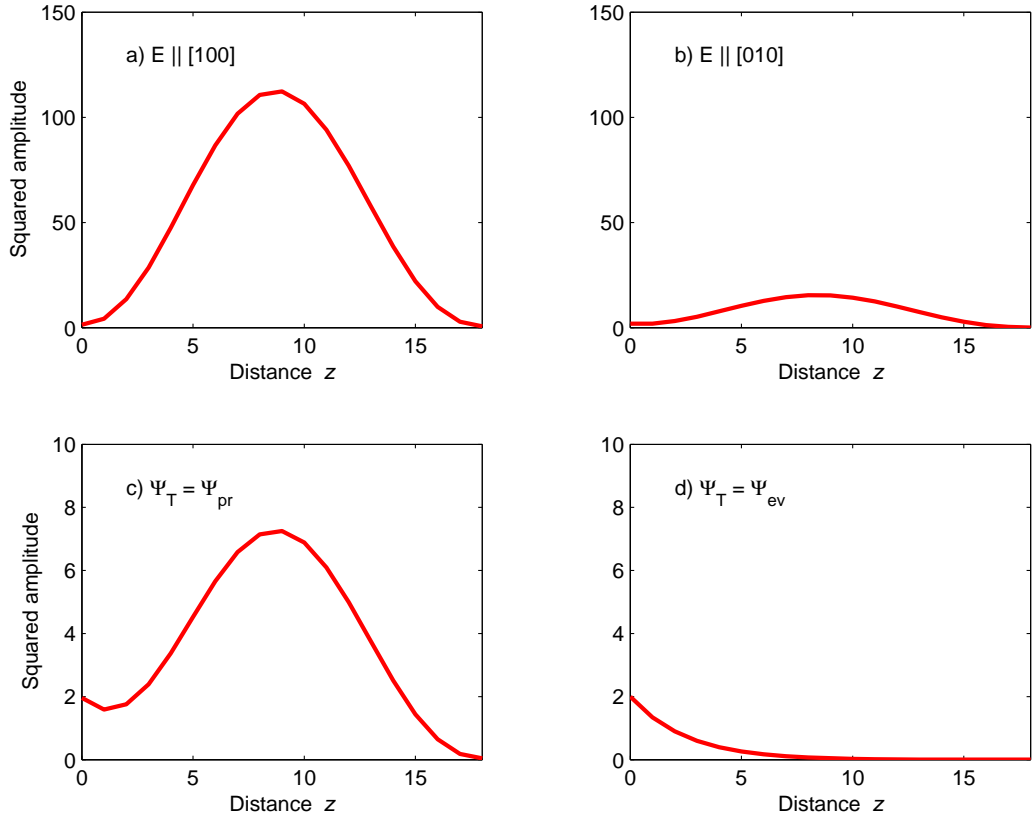}}
\caption{Smoothed energy density distribution $W\left( z\right) $ at
frequency of the first (closest to the DBE) giant transmission band edge
resonance in Fig. \protect\ref{tf_DBE} for four different polarizations of
the incident wave. In a single mode excitation regime of Figs. (c) and (d),
the transmission resonance is suppressed. Particularly so in the case (d),
where the EM field inside the periodic medium corresponds to a single
evanescent mode.}
\label{AMnad18DBE}
\end{figure}

\begin{figure}[tbph]
\scalebox{0.8}{\includegraphics[viewport=0 0 500 400,clip]{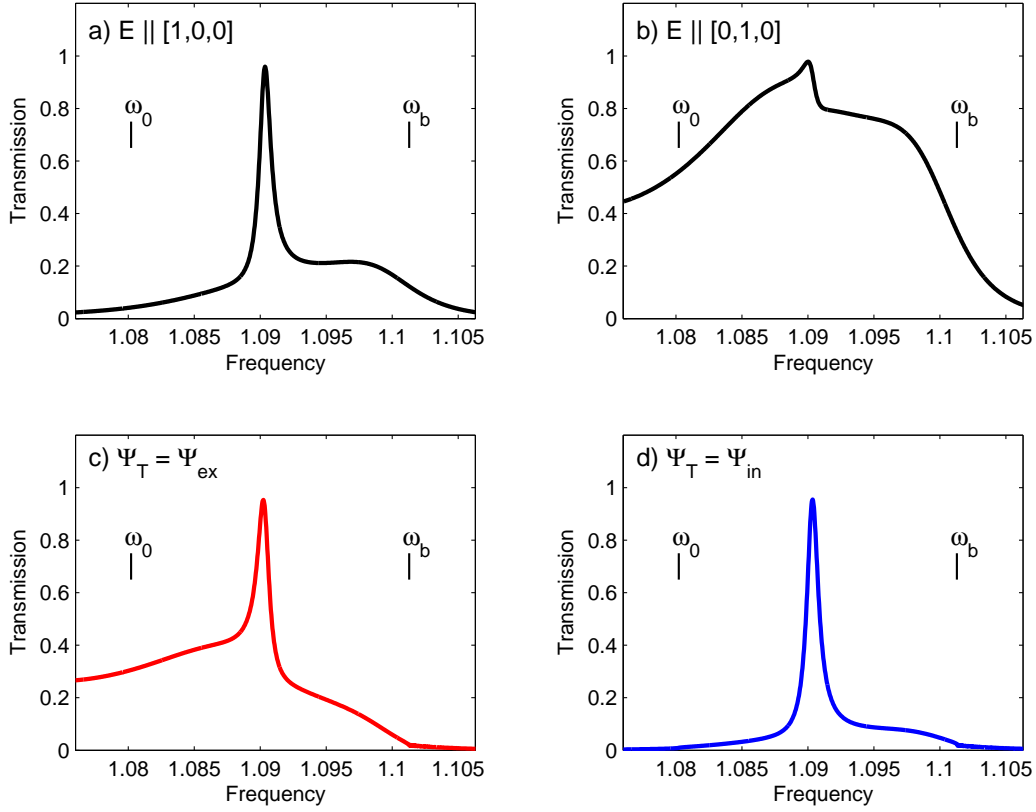}}
\caption{Manifestation of SBE related double resonance in the transmission
dispersion $t\left( \protect\omega\right) $ of periodic stack with $N=18$.
The respective $k-\protect\omega$ diagram is shown in Fig. \protect\ref{DR2}%
(b). Observe that at the resonance frequency, the stack transmittance is
close to unity regardless of the incident wave polarization. By contrast, in
the case of DBE-related giant transmission resonance in Fig. \protect\ref%
{tf_DBE}, the impedance matching is polarization dependent. In the cases (c)
and (d), the incident wave polarization is adjusted so that at any given
frequency it would correspond to the respective single mode excitation
regime in the semi-infinite layered structure.}
\label{tf18adSBE}
\end{figure}

Consider now the vicinity of a split photonic band edge (SBE) on the $%
k-\omega$ diagram in Fig. \ref{DR2}(b). The physical characteristics of the
periodic structure are chose so that the SBE in Fig. \ref{DR2}(b) is close
to a DBE. The frequency range%
\begin{equation}
\omega_{0}<\omega<\omega_{b},   \label{w_0<w<w_b}
\end{equation}
covers a narrow portion of the transmission band which includes the SBE. At
any given frequency from (\ref{w_0<w<w_b}), there are two pairs of
reciprocal Bloch waves with very low group velocity and different
polarizations; each pair being capable of producing its own slow-wave cavity
resonance with the resonance conditions similar to (\ref{k_m}) or (\ref%
{kappa_m, nu_m}). Our focus is on the possibility of the two resonances
occurring at the same frequency. Such a situation will be referred to as the 
\emph{double resonance}. It turns out that the double transmission band edge
resonance is as powerful as the giant resonance associated with a DBE. But,
in addition, a SBE related resonance utilizes all the energy of the incident
wave regardless of its polarization. By contrast, a DBE\ based giant
transmission resonance is coupled only with one polarization component of
the incident wave; the rest of the incident wave energy being reflected back
to space. This important difference is obvious if we compare the DBE related
transmission dispersion shown Fig. \ref{tf_DBE} and the SBE related
transmission dispersion shown in Fig. \ref{tf18adSBE}.

To start with, let us consider the dispersion curve with a SBE in more
detail. If the split between the twin band edges $b_{1}$ and $b_{2}$ in
Figs. \ref{DR2}(b) is small, the dispersion relation in the vicinity of SBE
can be approximated as%
\begin{equation}
\nu\left( \kappa\right) \approx\frac{a}{2}\kappa^{2}+\frac{b}{4}\kappa ^{4}, 
\label{DR SBE}
\end{equation}
where%
\begin{equation}
a/b<0   \label{a/b<0}
\end{equation}
and%
\begin{equation}
\ \left\vert a/b\right\vert \ll1.   \label{a/b<<1}
\end{equation}
The inequality (\ref{a/b<0}) is the condition for SBE. Indeed, in the
opposite case of%
\begin{equation}
a/b>0,   \label{a/b>0}
\end{equation}
the dispersion curve (\ref{DR SBE}) would develop a RBE at $\kappa=0$, as
shown in Fig. \ref{DR2}(a). While in the case%
\begin{equation}
a/b=0,   \label{a/b=0}
\end{equation}
the dispersion curve (\ref{DR SBE}) would have a DBE at $\kappa=0$. The
additional inequality (\ref{a/b<<1}) is the condition for the proximity of
the SBE to a DBE. This proximity allows us to use the expansion (\ref{DR SBE}%
) in the frequency range spanning both twin edges of the SBE. More
importantly, the condition (\ref{a/b<<1}) is essential for the phenomenon of
the giant transmission resonance in the vicinity of SBE.

There are three stationary points associated with a SBE. The first one is
trivial%
\begin{equation}
\kappa_{a}=0,\ \nu_{a}=0.   \label{k_a, w_a}
\end{equation}
It is located either at the center of the Brillouin zone, or at its
boundary. The other two stationary points correspond to the actual SBE%
\begin{equation}
\pm\kappa_{b}=\pm\sqrt{-a/b},\ \ \nu_{b}=-a^{2}/4b.   \label{k_b, w_b}
\end{equation}
Taking into account (\ref{k_b, w_b}), the condition (\ref{a/b<<1}) for the
proximity of the SBE to a DBE can be recast as%
\begin{equation}
\kappa_{b}\ll1.   \label{k_b<<1}
\end{equation}
The condition (\ref{k_b<<1}) implies that the points $b_{1}$ and $b_{2}$ on
the dispersion curve are close to each other.

In what follows, we assume for simplicity that%
\begin{equation}
b<0<a.   \label{b<0<a}
\end{equation}
In this case, the SBE in question corresponds to the upper edge of the
transmission band, as shown in Fig. \ref{tf_DBE}(b). The alternative case of%
\begin{equation}
a<0<b   \label{a<0<b}
\end{equation}
corresponds to the SBE\ being the lower edge of the respective transmission
band. There is no qualitative difference between the two cases.

At any given frequency $\nu$ within the range%
\begin{equation}
\nu_{a}<\nu<\nu_{b},   \label{w_a<w<w_b}
\end{equation}
there are two pairs of reciprocal Bloch waves. Each pair comprises one
forward and one backward propagating modes with equal and opposite wave
numbers and group velocities%
\begin{equation}
\pm\kappa_{in}=\pm\kappa_{b}\sqrt{1-\sqrt{1-\frac{\nu}{\nu_{b}}}},\ \ \nu
_{a}<\nu<\nu_{b}.   \label{k_in}
\end{equation}%
\begin{equation}
\pm\kappa_{ex}=\pm\kappa_{b}\sqrt{1+\sqrt{1-\frac{\nu}{\nu_{b}}}},\ \ \
\nu<\nu_{b},   \label{k_ex}
\end{equation}
The pair of wave numbers (\ref{k_in}) corresponds to the concave portion of
the dispersion curve (\ref{DR SBE}), while the pair of wave numbers (\ref%
{k_ex}) corresponds to the convex portion of the dispersion curve. Obviously,%
\begin{equation}
\kappa_{in}<\kappa_{ex},\text{ \ at \ }\ \nu_{a}<\nu<\nu_{b}. 
\label{k_in < k_ex}
\end{equation}
At any given frequency, the EM field inside the periodic stack is%
\begin{equation}
\Psi_{T}\left( z\right) =\Psi_{in}\left( z\right) +\Psi_{ex}\left( z\right)
,   \label{Psi(SBE)}
\end{equation}
where%
\begin{equation}
\Psi_{in}\left( z\right) =\Psi_{\kappa_{in}}\left( z\right) +\Psi
_{-\kappa_{in}}\left( z\right) ,   \label{Psi_in}
\end{equation}
and%
\begin{equation}
\Psi_{ex}\left( z\right) =\Psi_{\kappa_{ex}}\left( z\right) +\Psi
_{-\kappa_{ex}}\left( z\right) .   \label{Psi_ex}
\end{equation}
To distinguish between $\Psi_{in}\left( z\right) $ and $\Psi_{ex}\left(
z\right) $, we will refer to the respective quantities as the external and
internal. The inequality (\ref{k_in < k_ex}) justifies these terms.

\subsection{Conditions for the double SBE resonance}

\begin{figure}[tbph]
\scalebox{0.8}{\includegraphics[viewport=0 0 500 400,clip]{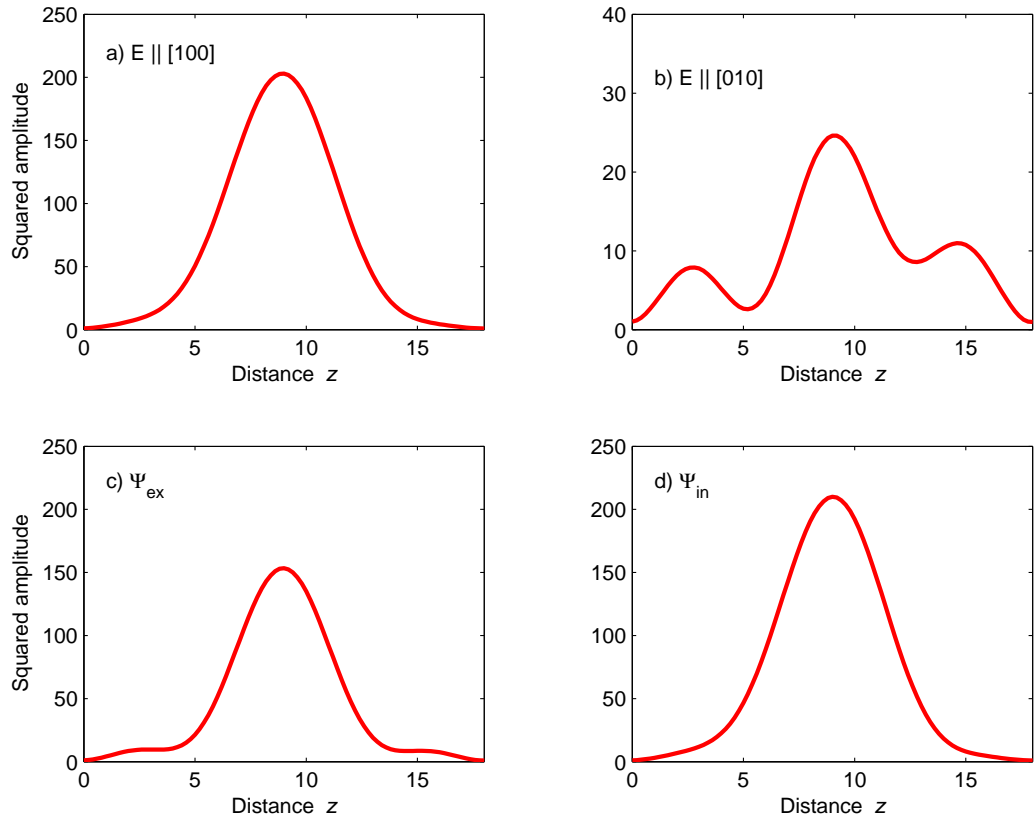}}
\caption{Smoothed energy density distribution at frequency of the SBE
related giant transmission resonance in Fig. \protect\ref{tf18adSBE} for
four different polarizations of the incident wave. The cases (c) and (d)
relate to a single mode excitation regime.}
\label{AMnad18SBE}
\end{figure}

\begin{figure}[tbph]
\scalebox{0.8}{\includegraphics[viewport=0 0 500 400,clip]{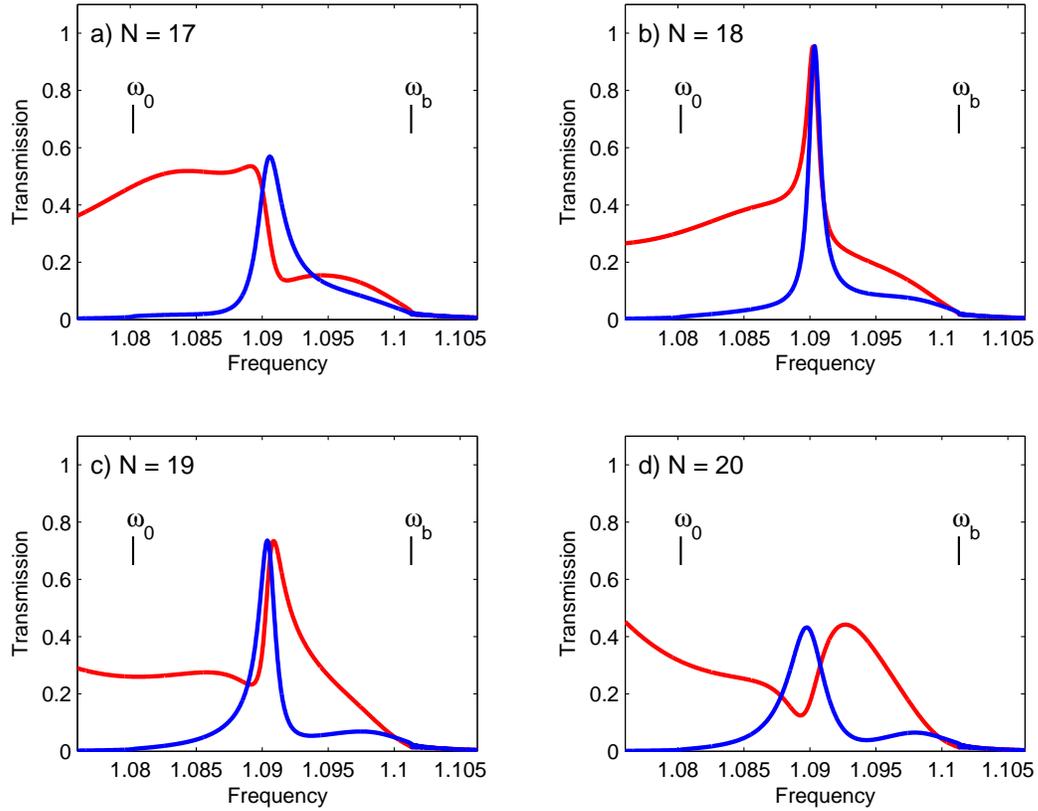}}
\caption{Transmission dispersion of periodic stacks composed of different
number $N$ of unit cells. The frequency range shown includes SBE on the $k-%
\protect\omega$ diagram in Fig. \protect\ref{DR2}(b). The red and the blue
curves correspond to two different polarizations of incident wave -- in
either case, the incident wave polarization is adjusted so that at any given
frequency it corresponds to a single mode excitation regime. In the case (b)
of $N=18$, the two resonance frequencies nearly coincide, creating condition
for double transmission resonance with perfect impedance matching.}
\label{tfNadSBE}
\end{figure}

Within the frequency range (\ref{w_a<w<w_b}), either pair of the reciprocal
Bloch waves (\ref{Psi_in}) and (\ref{Psi_ex}) can develop a transmission
resonance. Of particular interest here is the case where the two resonances
occurs at the same or almost the same frequency. This situation is
illustrated in Fig. \ref{tfNadSBE}(c), as well as in Fig. \ref{tf18adSBE},
and \ref{AMnad18SBE}.

Let us start with the internal resonance created by the reciprocal pair (\ref%
{Psi_in}) of Bloch waves corresponding to the concave section of the
dispersion curve. It is possible that the frequency range (\ref{w_a<w<w_b})
contains only a single cavity resonance -- the one with $m=1$. Such a case
is determined by either the proximity of the SBE to a DBE, or by the right
choice of the number $N$ of the unit cells in the stack. According to (\ref%
{k_b, w_b}) and (\ref{k_m}), the condition for a single internal resonance is%
\begin{equation}
\kappa_{1}<\kappa_{b}<2\kappa_{1},\ \ \kappa_{in}=\kappa_{1}=\frac{\pi}{N}. 
\label{k1<kb<2k1}
\end{equation}
The respective resonance frequency $\nu_{in}$ is determined by (\ref{w_m})
and (\ref{DR SBE})%
\begin{equation}
\nu_{in}=\nu_{1}=\frac{a}{2}\kappa_{1}^{2}+\frac{b}{4}\kappa_{1}^{4}, 
\label{w_in}
\end{equation}
where $\kappa_{1}=\pi/N$.

Consider now the external resonance created by the reciprocal pair (\ref%
{Psi_ex}) of Bloch waves corresponding to the convex section of the
dispersion curve. Let us impose the condition%
\begin{equation}
\nu_{ex}=\nu_{in}\equiv\nu_{r}   \label{w_ex=w_in}
\end{equation}
that the external and internal resonances occur at the same frequency (\ref%
{w_in}). This condition leads to the following equality%
\begin{equation*}
\ \frac{a}{2}\kappa_{in}^{2}+\frac{b}{4}\kappa_{in}^{4}=\frac{a}{2}\kappa
_{ex}^{2}+\frac{b}{4}\kappa_{ex}^{4}
\end{equation*}
Simple analysis shows that it is only possible if at $\nu=\nu_{r}$ we have%
\begin{equation*}
\kappa_{ex}=2\kappa_{in}=2\kappa_{1}=\frac{2\pi}{N}. 
\end{equation*}
The relation%
\begin{equation*}
\kappa_{ex}=2\kappa_{in}
\end{equation*}
together with (\ref{k_in}) and (\ref{k_ex}) yield%
\begin{equation*}
\kappa_{b}=\sqrt{\frac{5}{2}}\kappa_{1}=\sqrt{\frac{5}{2}}\frac{\pi}{N}. 
\end{equation*}
The frequency of the double transmission resonance is%
\begin{equation}
v_{r}=\left( \frac{4}{5}\right) ^{2}v_{b}=\frac{2}{5}b\left( \frac{\pi}{N}%
\right) ^{2}.   \label{v_r}
\end{equation}
The group velocities of the two reciprocal pairs of Bloch waves at the
resonance frequency $v_{r}$ are%
\begin{equation*}
u_{in}=\mp\frac{3}{2}b\kappa_{1}^{3}=\mp\frac{3}{2}b\left( \frac{\pi}{N}%
\right) ^{3}. 
\end{equation*}
and%
\begin{equation*}
u_{ex}=\pm3b\kappa_{1}^{3}=\pm3b\left( \frac{\pi}{N}\right) ^{3}. 
\end{equation*}

By comparison, in the case of a DBE related giant transmission band edge
resonance, we would have the following estimation for the resonance
frequency $v_{1}$ and the respective group velocity of the two Bloch
components%
\begin{equation*}
v_{1}\propto\frac{b}{N^{2}},\ u_{1}\propto\frac{b}{N^{3}}. 
\end{equation*}
These estimations are similar those related to the double transmission
resonance associated with SBE. In either case, the average resonance energy
density is estimated by (\ref{<W> RBE}), which justifies the term \emph{giant%
} transmission resonance.

Let us remark that the entire consideration of this subsection was based on
the assumption that each pair (\ref{Psi_in}) and (\ref{Psi_ex}) of the
reciprocal Bloch waves is responsible for its own individual transmission
resonance, described as a standing wave (\ref{Psi_in}) or (\ref{Psi_ex}),
respectively. While the double resonance at $\nu _{r}$ is described as the
situation where the frequencies $\nu _{in}$ and $\nu _{ex}$ of those
individual resonances merely coincide. In fact, these two transmission
resonances can be treated as independent only if the respective resonance
frequencies $\nu _{in}$ and $\nu _{ex}$ are well separated. As soon as $\nu
_{in}$ and $\nu _{ex}$ are close to each other, the EM field $\Psi
_{T}\left( z\right) $ in (\ref{Psi(SBE)}) becomes a superposition of all
four propagating Bloch modes. The latter situation persists even if the
incident wave polarization correspond to the, so-called, single mode
excitation regime, defined in the previous section. In other words, the
single mode excitation regime produces almost pure internal or external
resonance, provided that their frequencies are well separated. Otherwise,
the EM field $\Psi _{T}\left( z\right) $ is a superposition of all four
Bloch eigenmodes with two essentially different sets of wave numbers. In
such a case, the resonance field $\Psi _{T}\left( z\right) $ cannot be
viewed as a standing wave (\ref{1sw}). The physical reason for such a strong
hybridization is that due to the condition (\ref{k_b<<1}), the RBE in
question is very close to a DBE. On the other hand, in the vicinity of a
DBE, all four vector-columns $\Psi _{k}\left( z\right) $ in (\ref{4Psi_k})
become nearly parallel to each other \cite{PRE05,PRE06}. The latter
circumstance excludes the possibility of exciting only one of the two pairs
of the reciprocal Bloch modes (\ref{Psi_in}) or (\ref{Psi_ex}) in the
situation (\ref{w_ex=w_in}), where the resonance conditions (\ref{k_m}) are
in place for both of them simultaneously.  The strong hybridization implies
that the estimation (\ref{v_r}) of the double resonance frequency is rather
crude. Still, it provides a very useful guidance on the conditions for SBE
related giant transmission resonance. The bottom line is that the SBE
related giant transmission resonance is as powerful as that related to a
DBE. But, in addition, the SBE related resonance provides a perfect coupling
with the incident wave regardless of its polarization.

\section{Conclusion}

In summary, we would like to stress that the remarkable features of the DBE
and SBE related giant transmission resonances can be derived from such
fundamental characteristics of the periodic composite medium as its
electromagnetic dispersion relation. Specific details of the periodic array,
such as physical characteristics of the constitutive components, or
structural geometry, are only important as long as the symmetry of the
periodic array is compatible with the existence of the required spectral
singularities.\bigskip

\textbf{Acknowledgment and Disclaimer:} Effort of A. Figotin and I.
Vitebskiy is sponsored by the Air Force Office of Scientific Research, Air
Force Materials Command, USAF, under grant number FA9550-04-1-0359.\ The
authors are thankful to A. Chabanov for stimulating discussions.

\bigskip\pagebreak

\bigskip

\bigskip\pagebreak

\bigskip

\section{Appendix}

\begin{figure}[tbph]
\scalebox{0.8}{\includegraphics[viewport=-100 0 500 250,clip]{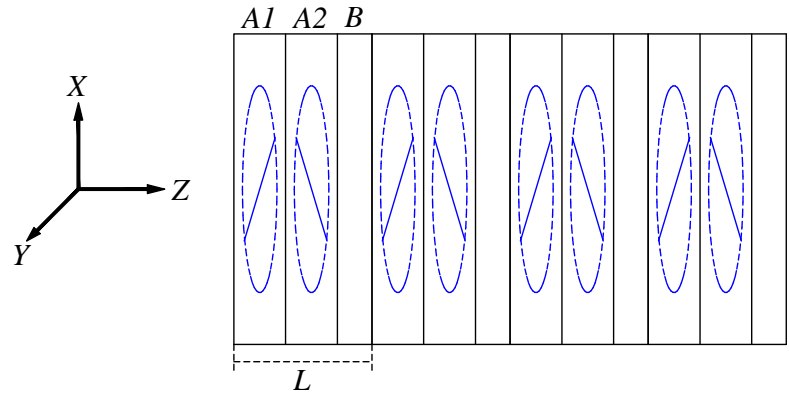}}
\caption{Periodic layered structure with a unit cell $L$ containing two
misaligned anisotropic $A$ layers, and one isotropic $B$ layer. The
respective dielectric permittivity tensors are given in (\protect\ref{eps B}%
), (\protect\ref{eps A(phi)}), and (\protect\ref{eps NUM}). This is the
simplest layered array supporting \ the Bloch dispersion relation with a DBE
and/or a SBE, as shown in Fig. \protect\ref{DR2}(b).}
\label{StackAAB}
\end{figure}

The simplest periodic layered structure supporting a DBE or a SBE at normal
propagation is shown in Fig. \ref{StackAAB}. A unit cell $L$ contains one
isotropic $B$ layer and two misaligned anisotropic layers $A_{1}$ and $A_{2}$
with inplane anisotropy. The isotropic layers have the thickness $B$ and the
dielectric permittivity%
\begin{equation}
\hat{\varepsilon}_{B}=\left[ 
\begin{array}{ccc}
\varepsilon_{B} & 0 & 0 \\ 
0 & \varepsilon_{B} & 0 \\ 
0 & 0 & \varepsilon_{B}%
\end{array}
\right] .   \label{eps B}
\end{equation}
The dielectric permittivity tensors $\hat{\varepsilon}_{A}$ in each
anisotropic $A$ layer has the form%
\begin{equation}
\hat{\varepsilon}_{A}\left( \varphi\right) =\left[ 
\begin{array}{ccc}
\varepsilon_{A}+\delta\cos2\varphi & \delta\sin2\varphi & 0 \\ 
\delta\sin2\varphi & \varepsilon_{A}-\delta\cos2\varphi & 0 \\ 
0 & 0 & \varepsilon_{3}%
\end{array}
\right] ,   \label{eps A(phi)}
\end{equation}
where the parameter $\delta$ characterizes the magnitude of inplane
anisotropy, and the angle $\varphi$ determines the orientation of the
anisotropy axes in the $x-y$ plane. All the $A$ layers have the same
thickness $A$ and the same magnitude $\delta$ of inplane anisotropy. The
only difference between the adjacent anisotropic layers $A_{1}$ and $A_{2}$
in \ref{StackAAB} is their orientation $\varphi$. An important
characteristic of the periodic structure in Fig. \ref{StackAAB} is the
misalignment angle%
\begin{equation}
\phi=\varphi_{1}-\varphi_{2}   \label{phi}
\end{equation}
between the layers $A_{1}$ and $A_{2}$. This angle determines the symmetry
of the periodic array and, eventually, the kind of $k-\omega$ diagram it can
display.

In all numerical simulations related to the periodic layered structure in
Fig. \ref{StackAAB} we use the following values of the material parameters
in (\ref{eps B}), (\ref{eps A(phi)}) and (\ref{phi})%
\begin{equation}
\ \varepsilon_{B}=16.0,\varepsilon_{A}=4.7797,\ \delta=3.4572,\phi=\pi/6. 
\label{eps NUM}
\end{equation}
At normal propagation, the numerical value of $\varepsilon_{3}$ in (\ref{eps
A(phi)}) is irrelevant. The relative thickness of the $A$ and $B$ layers,
can be different in different examples.

In all plots of the field distribution inside periodic media at $0<z<D$ we,
in fact, plotted the following physical quantity%
\begin{equation}
\left\langle \left\vert \Psi\left( z\right) \right\vert ^{2}\right\rangle
=\left\langle \vec{E}\left( z\right) \cdot\vec{E}^{\ast}\left( z\right) +%
\vec{H}\left( z\right) \cdot\vec{H}^{\ast}\left( z\right) \right\rangle
_{L},   \label{Sm Int}
\end{equation}
which is the squared field amplitude averaged over a local unit cell. The
actual function $\left\vert \Psi\left( z\right) \right\vert ^{2}$, as well
as the electromagnetic energy density distribution $W\left( z\right) $, are
strongly oscillating functions of the coordinate $z$, with the period of
oscillations coinciding with the unit cell length $L$. Given the relation $%
W\propto\left\vert \Psi\left( z\right) \right\vert ^{2}$, the quantity (\ref%
{Sm Int}) can also be qualitatively interpreted as the smoothed energy
density distribution, with the correction coefficient of the order of unity.

In all plots, the distance $z,$ the wave number $k$, and the frequency $%
\omega$ are expressed in units of $L$, $L^{-1}$, and $cL^{-1}$, respectively.

\end{document}